\documentclass{kluwer}    
\usepackage{graphics}

\newdisplay{guess}{Conjecture}

\newcommand{\teff}{T$_{\mathrm{eff}}$}
\newcommand{\logg}{$\log{g}$}

\begin{document}                                                                                   
\begin{article}
\begin{opening}         
\title{A critical appraisal of ATLAS9 and NextGen 5 model atmospheres} 
\author{Emanuele \surname{Bertone}$^{1,2}$}
\author{Alberto \surname{Buzzoni}$^{1,3}$}
\author{Miguel \surname{Ch\'avez}$^{4}$}
\author{Lino H. \surname{Rodr\'\i guez}$^{4}$}

\runningauthor{E. Bertone et al.}
\runningtitle{}

\institute{$^1$Osservatorio Astronomico di Brera, Italy, $^2$Universit\`a
Statale di Milano, Italy, $^3$Telescopio Nazionale Galileo, Spain, $^4$INAOE, Mexico.}

\begin{abstract}
The fitting atmosphere parameters (\teff, g, and [Fe/H]) for over 300 stars in
the Gunn \& Striker (1983) and Jacoby {\it et al.}\ (1984) catalogs have been obtained
relying on the Kurucz (1992) ATLAS9 and Hauschildt {\it et al.}\ (1999) NextGen5 synthesis
models. The output results are compared, and a critical appraisal of both
theoretical  codes is performed.
\end{abstract}
\end{opening}           

As a major improvement over the standard ATLAS9 code for model atmospheres by Kurucz (1992),
the new NextGen5 synthesis code of Hauschildt {\it et al.}\ (1999) adopts a more refined 
treatment of molecular opacity, and includes spherical simmetry in the atmosphere layers 
for low-gravity models. In order to assess the main differences between the two theoretical 
codes, it is of special interest to investigate the theoretical temperature calibration for 
stars of different spectral type.

In this sense, we devised a procedure to determine the fundamental
parameters of a star (\teff, \logg, [M/H]) by comparing its observed 
spectral energy distribution with a grid of synthetic spectra. The fiducial 
best model is identified by minimizing the $\Delta \log ({\rm flux})$ 
standard deviation, $\sigma(f)$, over the full wavelength range of the observations.

An application to  the Gunn \& Stryker (1983, hereafter ``GS'') and Jacoby {\it et al.}\ (1984, 
``JHC'') atlases provided the fitting parameters for over 300 stars
by matching with the solar metallicity model grids of ATLAS9 
($3500 \le T_\mathrm{eff} \le 50\,000$~K, $0.0 \le \log{g} \le 5.0$) and 
NextGen5 ($2000 \le T_\mathrm{eff} \le 10\,000$~K, $0.0 \le \log{g} \le 5.5$).

\begin{figure}
\resizebox{!}{5cm}{{\includegraphics{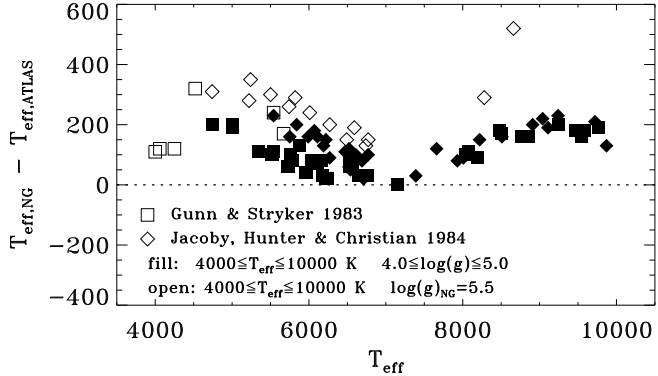}}}
\caption{The difference in the fitting value of $T_\mathrm{eff}$ for the GS and JHC
stars as derived from the ATLAS9 and Nextgen5 best-fit models.}
\end{figure}

As shown in Figure 1, when comparing the fiducial temperature for GS and JHC class V stars, 
the NextGen5 fit results in a significantly warmer value of \teff\ with respect to ATLAS9,
especially at the extreme edges of the \teff\ range.

\begin{figure}
\resizebox{5.3cm}{5cm}{{\includegraphics{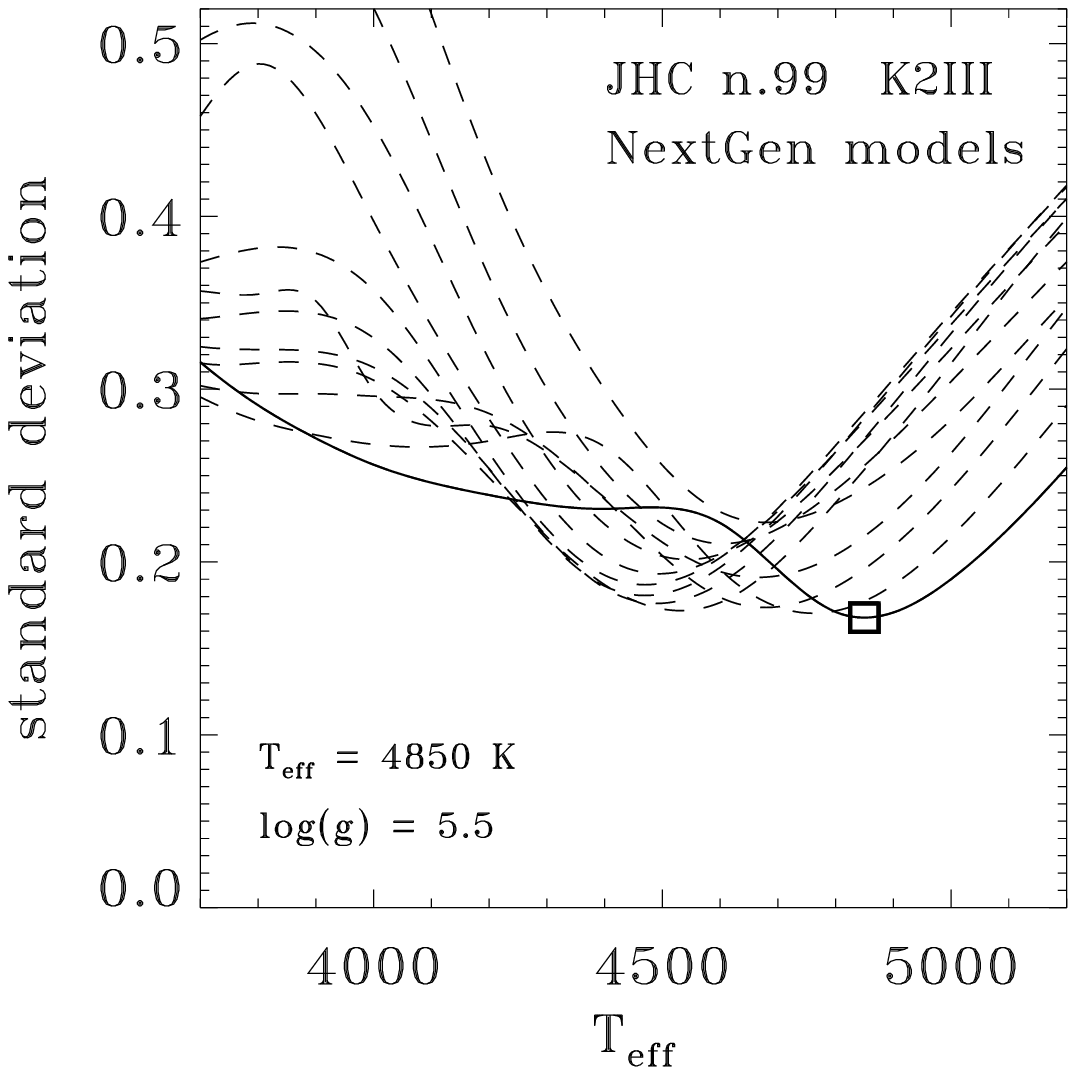}}}
\resizebox{5.3cm}{5cm}{{\includegraphics{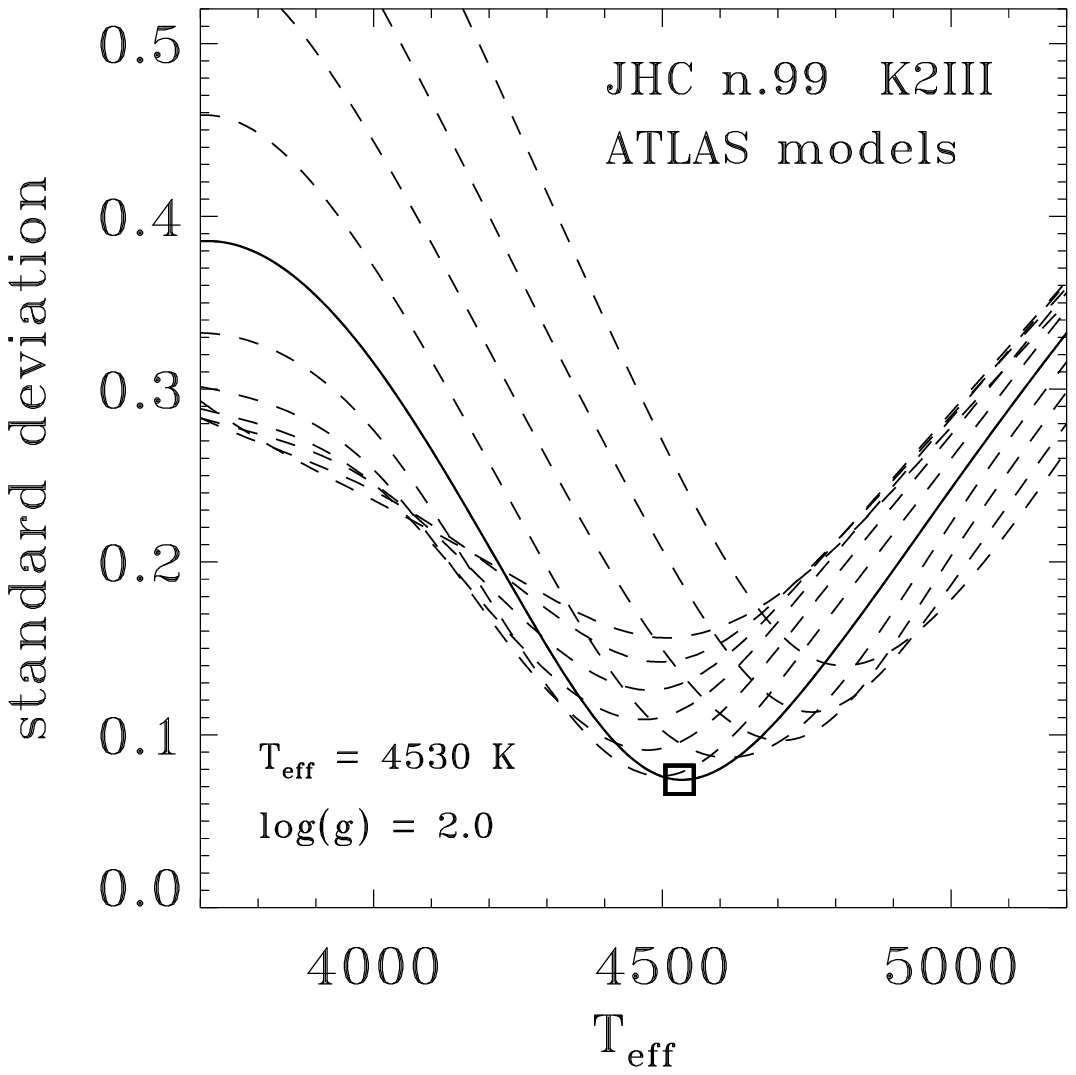}}}
\caption{An example of the fitting procedure for star no.\ 99 in the JHC sample.
For ATLAS9 and Nextgen5 models, the open square marks the optimum fit among the
different $\sigma (f)$ equi-gravity envelope curves. A solar metallicity is assumed.}
\end{figure}

Figure~2 is an example of our best-fit procedure for a star in the JHC atlas 
comparing with both ATLAS9 and NextGen5 models. Equi-gravity envelope curves
for the standard deviation $\sigma (f)$ across the temperature range are reported
searching for the absolute minimum that marks the fitting \teff\ and \logg.
A more univocal and ``sharper'' solution is in general reached by the ATLAS9 fit, 
while NextGen5 models display a more entangled trend for the $\sigma (f)$ function,
especially as far as cool stars (Sp. type K and M) are concerned.

\end{article}
\end{document}